# Correlated interlayer exciton insulator in double layers of monolayer WSe$_2$ and moiré WS$_2$/WSe$_2$

Zuocheng Zhang[1], Emma C. Regan[1,2,3], Danqing Wang[1,2,3], Wenyu Zhao[1], Shaoxin Wang[1], Mohammed Sayyad[4], Kentaro Yumigeta[4], Kenji Watanabe[5], Takashi Taniguchi[6], Sefaattin Tongay[4], Michael Crommie[1,3,7], Alex Zettl[1,3,7], Michael P. Zaletel[1,3], Feng Wang[1,3,7]∗

[1] Department of Physics, University of California at Berkeley, Berkeley, California 94720, United States

[2] Graduate Group in Applied Science and Technology, University of California at Berkeley, Berkeley, California 94720, United States

[3] Material Science Division, Lawrence Berkeley National Laboratory, Berkeley, California 94720, United States

[4] School for Engineering of Matter, Transport and Energy, Arizona State University, Tempe, Arizona 85287, United States

[5] Research Center for Functional Materials, National Institute for Materials Science, 1-1 Namiki, Tsukuba 305-0044, Japan

[6] International Center for Materials Nanoarchitectonics, National Institute for Materials Science, 1-1 Namiki, Tsukuba 305-0044, Japan

[7] Kavli Energy NanoSciences Institute at University of California Berkeley and Lawrence Berkeley National Laboratory, Berkeley, California 94720, United States

∗ Correspondence to: fengwang76@berkeley.edu

**Abstract**:

Moiré superlattices in van der Waals heterostructures have emerged as a powerful tool for engineering novel quantum phenomena. Here we report the observation of a correlated interlayer exciton insulator in a double-layer heterostructure composed of a $WSe_2$ monolayer and a $WS_2/WSe_2$ moiré bilayer that are separated by an ultrathin hexagonal boron nitride (hBN). The moiré $WS_2/WSe_2$ bilayer features a Mott insulator state at hole density $p/p_0 = 1$, where $p_0$ corresponds to one hole per moiré lattice site. When electrons are added to the Mott insulator in the $WS_2/WSe_2$ moiré bilayer and an equal number of holes are injected into the $WSe_2$ monolayer, a new interlayer exciton insulator emerges with the holes in the $WSe_2$ monolayer and the electrons in the doped Mott insulator bound together through interlayer Coulomb interactions. The excitonic insulator is stable up to a critical hole density of ~ $0.5p_0$ in the $WSe_2$ monolayer, beyond which the system becomes metallic. Our study highlights the opportunities for realizing novel quantum phases in double-layer moiré systems due to the interplay between the moiré flat band and strong interlayer electron interactions.

**Main text**:

Excitonic insulators form when electrons and holes bind into pairs through an attractive Coulomb interaction[1,2]. The realization of excitonic insulators has been actively pursued for many decades[3]. Most striking observations of excitonic insulators are demonstrated in quantum Hall double layers[4–7], where Landau levels in a strong magnetic field are flat electronic bands that suppress the kinetic energy and enhance the electron-hole correlation. Flat electronic bands can also be achieved in moiré superlattices, which has enabled the observation of correlated insulators[8–14], superconductivity[9,11,14–16], Chern insulators[17–20], moiré excitons[21–24] and generalized Wigner crystal states[13,25] in different moiré heterostructures. By integrating the moiré superlattice into a double-layer system, where the moiré superlattice is separated from another semiconductor layer by an ultrathin hexagonal boron nitride (hBN) layer, we can simultaneously achieve a flat electronic band and strong interlayer electron-hole coupling. This interplay for moiré flat bands and strong interlayer electron-hole interactions can lead to novel correlated quantum phases, including excitonic insulators at zero magnetic fields.

Here we demonstrate a new type of correlated interlayer exciton insulator phases in a double-layer heterostructure composed of a $WS_2/WSe_2$ moiré bilayer strongly coupled to a $WSe_2$ monolayer that is separated by a 1nm thick hBN. Charge neutral $WSe_2$ monolayer has been used as a sensor to probe nearby Wigner crystal insulators in the $WS_2/WSe_2$ heterostructures in reference 25. Our study explores the double-layer system where both the $WSe_2$ monolayer and the moiré heterostructure are doped with charge carriers that interact strongly with each other (Fig. 1a). The interlayer electron correlation leads to the formation of a new interlayer exciton insulator state composed of holes of a band insulator (in the $WSe_2$ monolayer) and electrons of a

Mott insulator (in the WS$_2$/WSe$_2$ moiré bilayer). We utilize the delicate dependence of the 2s exciton resonance on the dielectric environment to determine the exciton insulating phase of the double-layer heterostructure [25]. At the same time, we track the charge distributions in the WSe$_2$ monolayer and the moiré bilayer using the trion state in the WSe$_2$ monolayer and the interlayer moiré exciton photoluminescence (PL) in the WS$_2$/WSe$_2$ moiré bilayer, respectively.

Figure 1a shows a schematic of a double-layer heterostructure device with both the top ($V_t$) and bottom ($V_b$) gates. The WS$_2$/WSe$_2$ moiré bilayer has a near zero-twist-angle or 60-degrees-twist-angle. It features flat moiré bands with strong on-site and long-range Coulomb interactions, which results in a prominent Mott insulator at hole density $p/p_0 = 1$ and generalized Wigner crystal states at $p/p_0 = 2/3$ and $1/3$ (ref. 13). Here $p_0$ corresponds to one hole per moiré lattice site. We chose the WSe$_2$ monolayer as the second layer because it has a large effective mass and strong electron correlations[26]. These two layers are separated by an ultrathin hBN layer with approximately 1nm thickness. Thicker hBN layers are used for the gate dielectric layers. The electrical contacts to the transition metal dichalcogenide (TMDC) flakes and the top/bottom gates are made of few-layer graphite (FLG). The stack is further capped with another hBN layer. The dual gate configuration enables us to independently control the total hole concentration $p = -\frac{1}{e}[\frac{\varepsilon_{hBN}\varepsilon_0}{d_t}(V_t - V_{t0}) + \frac{\varepsilon_{hBN}\varepsilon_0}{d_b}(V_b - V_{b0})]$ and the vertical electric field $E = (\frac{1}{d_b}V_b - \frac{1}{d_t}V_t)$, where $V_{t0}$ and $V_{b0}$ are the onset gate voltages to inject holes into the double layers, $d_b$ ($d_t$) is the thickness of the bottom (top) hBN dielectric layer, and $\varepsilon_{hBN}$ is dielectric constant of hBN with $\varepsilon_{hBN} \approx 4.2$ (ref.13). The vertical electric field can change the relative potential of the WSe$_2$ monolayer and the WS$_2$/WSe$_2$ moiré bilayer, and consequently, the distribution of the hole concentration in the WSe$_2$ monolayer ($p_{mono}$) and WS$_2$/WSe$_2$ moiré bilayer ($p_{moiré}$). Figure 1b illustrates the band alignment of the three TMDC layers in the heterostructure. The conduction band edge of the $WS_2$ monolayer has much lower energy than that in $WSe_2$ monolayers. As a result, electrostatically doped electrons tend to be confined in the moiré superlattice $WS_2$ layer. On the other hand, holes can be distributed in either the $WSe_2$ monolayer or the moiré superlattice $WSe_2$ layer depending on the vertical electric field: a positive $E$ will shift holes to the moiré superlattice $WSe_2$ layer while a negative $E$ will prefer holes in the $WSe_2$ monolayer.

We utilize different excitonic features in the optical spectra to selectively probe the electron and hole doping in individual TMDC layers and the correlated insulator states. Three different devices have been studied (Fig. S1), and they all exhibit similar behavior. We will focus on device I in the main text. (See supplementary information for experimental data of device II and III.) Figure 1c shows a broadband reflection contrast spectrum of the device with both the monolayer $WSe_2$ and the $WS_2/WSe_2$ moiré bilayer at charge neutral. (All experimental measurements are performed at a nominal temperature of $T = 1.6$ K unless otherwise specified.) The two absorption peaks centered around 1.695 eV and 1.790 eV are moiré exciton resonances of the $WS_2/WSe_2$ moiré superlattice[23]. The strongest absorption peak, around 1.722 eV, corresponds to the 1s exciton transition of the $WSe_2$ monolayer, which coincides with an additional weak moiré exciton state[23]. The other sharp transition around 1.847 eV is from the 2s exciton transition of the $WSe_2$ monolayer. The 1s exciton resonance of the $WSe_2$ monolayer has a single peak with a narrow line width comparable to that of isolated monolayers.

We investigate the correlated interlayer electronic states by varying both total hole densities and the vertical $E$ field in the double-layer system. Figure 2a to 2d, show the two-dimensional color plot of derivative reflection contrast spectra and PL spectra in three different photon energy windows (row I, II, and III) at the $E$ field of 160 mV/nm (column a), 16 mV/nm (column b), 0 mV/nm (column c), and -16 mV/nm (column d), respectively. The upper panels (row I) plot the first derivative spectra of the WSe$_2$ monolayer 2s exciton resonance ($d(-\Delta R/R)/dE_{photon}$) with respect to photon energy between 1.810 eV and 1870 eV, which probes the correlated insulating state in the double-layer system due to its sensitivity to the dielectric screening in the system[25]. The middle panels (row II) show the exciton and trion PL spectra of the WSe$_2$ monolayer between 1.680 eV to 1.735 eV, which probe the hole density in the WSe$_2$ monolayer. The lower panels (row III) display the interlayer exciton (IX) PL spectra of the WS$_2$/WSe$_2$ moiré bilayer between 1.318 eV to 1.550 eV, which probes the hole density in the moiré bilayer.

Figure 2aI shows the monolayer WSe$_2$ 2s exciton derivative spectra as a function of the total hole concentration ($p/p_0$) at $E$ = 160 mV/nm, where $p_0$ is estimated to be $(1.97 \pm 0.20) \times 10^{12}/cm^2$ (Fig. S2). The large positive $E$ field confines doped holes in the moiré bilayer and keeps the WSe$_2$ monolayer charge neutral. Well-defined but red-shifted monolayer WSe$_2$ 2s exciton resonances (arrows in Fig. 2aI) can be observed at discrete doping levels of $p/p_0$ = 1/3, 2/3, and 1. These 2s exciton resonances result from reduced free-carrier screening and are signatures of correlated insulating states[25]. Since only the moiré heterostructure is doped with carriers at $E$ = 160 mV/nm for $p/p_0$ < 2, these correlated insulating states can be attributed to the known generalized Wigner crystal states at $p/p_0$ = 1/3 and 2/3 and the Mott insulator state at $p/p_0$ = 1 (ref.13,25). Figure 2aII displays the PL spectra at the WSe$_2$ monolayer 1s exciton and trion transition energies. The 1s exciton PL is unchanged, and no trion PL is observed for $p/p_0 < 2$. They confirm that the hole density is zero in the WSe$_2$ monolayer. Figure 2aIII displays the IX PL spectra of the WS$_2$/WSe$_2$ moiré bilayer. The IX PL shows a discrete change at $p/p_0 = 1$, where the PL resonance energy blueshifts suddenly and the PL intensity exhibits a sudden increase. This abrupt increase of IX PL intensity provides a reliable signature of the Mott insulating state with one hole at each moiré superlattice ($p_{\text{moiré}}/p_0 = 1$). It coincides with $p/p_0 = 1$ in Fig. 2a because all the doped holes are in the moiré bilayer with $p_{\text{moiré}} = p$ in this case.

Figure 2bI shows the monolayer WSe$_2$ 2s exciton derivative spectra as a function of the total hole doping ($p/p_0$) at $E = 16$ mV/nm. Again, a prominent 2s exciton resonance is observed at $p/p_0 = 1$, indicating a correlated insulator state when the total doping is at one hole per moiré superlattice site. Figure 2bII displays the 1s exciton and trion PL spectra of the WSe$_2$ monolayer. Clear trion PL can be observed at large $p/p_0$, accompanied by the suppression of 1s exciton PL. It shows unambiguously that holes are now doped into the WSe$_2$ monolayer. In particular, the trion PL is present at $p = p_0$, indicating that the correlated insulator at $E = 16$ mV/nm already has partial hole doping in the WSe$_2$ monolayer. Figure 2bIII displays the IX PL spectra of the moiré bilayer. We find that the sudden change in the IX PL now takes place at doping $p/p_0 > 1$. It shows that a higher total hole density ($p/p_0 > 1$) is required to realize a filled moiré superlattice (i.e., $p_{\text{moiré}}/p_0 = 1$), confirming that holes are now distributed in both the WSe$_2$ monolayer and the WS$_2$/WSe$_2$ moiré bilayer at $p/p_0 = 1$.

Figure 2c and 2d, show corresponding data at $E = 0$ mV/nm and $E = -16$mV/nm, respectively. The correlated insulator state can be observed in both Fig. 2cI and 2dI at total doping $p/p_0 = 1$, although the corresponding 2s resonance becomes relatively weak at $E = -16$ mV/nm in Fig. 2dI. Significant hole doping of WSe$_2$ monolayer is present in the $p/p_0 = 1$ correlated insulator state at $E = 0$ mV/nm, as reflected in the appreciable WSe$_2$ trion PL signal at the dashed line of Fig. 2cII. The hole doping in the WSe$_2$ monolayer becomes even stronger at $E = -16$mV/nm (Fig. 2dII). At the same time, the IX PL spectra in Fig. 2cIII and 2dIII shows the hole doping in the moiré bilayer is much lower than the total hole density: the moiré bilayer Mott insulator state ($p_{moiré}/p_0 = 1$), defined by the sudden change of the IX PL resonance, occurs at a total hole doping $p$ much larger than $p_0$. At the correlated insulator state of $p/p_0 = 1$ (dashed lines in Fig. 2cIII and 2dIII), $p_{moiré}$ is smaller than $p_0$. They provide independent evidence that both the WSe$_2$ monolayer and the WS$_2$/WSe$_2$ moiré bilayer are appreciably hole-doped when the full double-layer system is in the correlated insulator state ($p/p_0 = 1$) at $E = 0$ mV/nm and $E = -16$ mV/nm.

We map out the phase diagram of the new correlated insulating state at $p/p_0 = 1$ and different $E$ in Fig. 3. Figure 3a shows the integrated trion area (probe of the WSe$_2$ monolayer hole density) as a function of total hole concentration $p/p_0$ and vertical electric field $E$. The corresponding data of the derivative of IX PL intensity ($dI_{IX}/d(p/p_0)$) (probe of the moiré bilayer hole density) and the 2s exciton signal (probe of the correlated insulator state) are displayed in Fig. 3b and 3c, respectively. The WSe$_2$ monolayer trion signal in Fig. 3a reveals two distinct regions for hole doping in the WSe$_2$ monolayer: the WSe$_2$ monolayer is charge neutral without any trion PL signal in region I (when $E$ field is below the black dashed line), and it becomes hole-doped with finite trion PL signal in region II (when $E$ field is above the black dashed line). Figure 3b provides an independent determination of the Mott insulator state ($p_{moiré}/p_0 = 1$) in the moiré bilayer, as characterized by the sudden increase of IX PL signal (i.e., maximum of $dI_{IX}/d(p/p_0)$ denoted by the green dashed line). The Mott insulator state ($p_{moiré}/p_0 = 1$) coincides with the vertical line defining $p/p_0 = 1$ in region I, as expected when the WSe$_2$ monolayer is charge neutral and all the doped holes are in the moiré bilayer. However, in region II, the $p_{moiré}/p_0 = 1$ dashed line has a finite slope and persists to the highest doping density and negative $E$ field. An increasingly higher total hole density $p/p_0$ is required to sustain the Mott insulator state ($p_{moiré}/p_0 = 1$) at more negative $E$. It confirms that an increasingly larger portion of holes is doped into the WSe$_2$ monolayer at positions deeper into region II. The WSe$_2$ monolayer 2s exciton signal in Fig. 3c probes the correlated insulating states of double layers. It shows that the correlated insulator state at $p/p_0 = 1$ can be stable in both region I and region II. The correlated insulator in region I, where $p_{mono} = 0$, is defined by the Mott insulator phase in the WS$_2$/WSe$_2$ moiré heterostructure. This insulator state extends into region II with the combined hole density defined by $p/p_0 = (p_{moiré} + p_{mono})/p_0 = 1$, and they eventually disappear at a large negative electric field (i.e., sufficiently high $p_{mono}$). Similar phase diagrams are observed in all three devices (Fig. S3).

Next, we compare the hole distribution at points A and B of Fig. 3c, where the total hole density is at $p/p_0 = 1$. Figure 3d illustrates the hole distribution at point A, where the holes fill each moiré lattice site in the WS$_2$/WSe$_2$ moiré layer to form a Mott insulator. Figure 3e illustrates the hole distribution at point B, where finite hole density is present in both the top moiré bilayer and the bottom WSe$_2$ monolayer. The inhomogeneity in the hole doping in both the WSe$_2$ monolayer and the moiré bilayer is relatively small, as shown by the rather narrow phase boundary of the moiré bilayer Mott insulating state in Fig. 3b and the correlated insulator state in Fig. 3c. Therefore, the correlated insulator with distributed holes in both the $WSe_2$ monolayer and the moiré bilayer should be a new quantum phase that is relatively homogeneous. This new correlated interlayer insulator has total doping of "one hole per moiré lattice site", but some of the holes are present in the $WSe_2$ monolayer that do not experience the moiré potential directly. This interlayer insulator is stabilized by the strong interlayer Coulomb interaction. Because the moiré bilayer to monolayer distance of $d \sim 1$nm is much smaller than the moiré length scale $L \sim 8$ nm, the interlayer Coulomb interaction can be extremely strong. It prevents the holes in the $WSe_2$ monolayer from occupying positions where the $WS_2/WSe_2$ moiré lattice site above already has a hole.

Figure 3f illustrates another perspective of this interlayer insulator by applying a particle-hole transformation relative to the Mott insulator state in the moiré bilayer. It clearly shows that electrons doped into the Mott insulator in the moiré bilayer can spontaneously bind the holes doped into the $WSe_2$ monolayer and form tightly bound interlayer excitons. Consequently, the insulating state at point B is described by an interlayer exciton insulator phase, where the interlayer exciton density $n_X$ is the same as $p_{mono}$. This interlayer exciton insulator phase is stable up to an exciton density of $n_X \sim 0.50 p_0$ (see supplementary information for details). When $n_X$ is further increased, the interlayer excitons start to dissociate, and the interlayer exciton insulator dissolves into a metallic phase.

Our phase diagram in Fig. 3c shows that weaker insulating states are also present at $p/p_0 = 2/3$ and $1/3$. These states are known as the generalized Wigner crystal states in region I. They also extend to a finite phase space in region II and can be understood as interlayer exciton insulators relative to the respective generalized Wigner crystal states in the moiré bilayer.

These correlated interlayer exciton insulators are quite unusual because they are composed of electrons in a Mott insulator (or generalized Wigner crystal insulator) and holes in a band insulator. Because electrons in the moiré bilayer may exhibit nontrivial spin configurations at the Mott insulator state, the nature of the interlayer exciton insulator can have interesting behavior associated with the magnetic order. In the simplest scenario, however, we can ignore the interlayer exciton's spin by considering temperatures above the Heisenberg scale to focus on the charge degree of freedom. In this case, the excitons are bosons and can potentially form an exciton condensate at sufficiently low temperatures. A macroscopic two-dimensional exciton superfluid can be realized at the Berezinskii-Kosterlitz-Thouless (BKT) transition with $n_x^{BKT}$ defined by $n_x^{BKT} = \frac{m_x k_B T}{1.3\, \hbar^2}$ (ref. 27,28). Here $\hbar$ is the reduced Planck constant, $m_x$ is the exciton mass $m_x = m_e + m_h$, $m_e$ and $m_h$ are the mass of electrons and holes, $k_B$ is the Boltzmann constant, and $T$ is the temperature. Figure 3g illustrates the different correlated phases in the double layers as a function of $p_{mono}$ at a fixed $p/p_0 = 1$. We start from the Mott insulator phase when $p_{mono} = 0$. The Mott insulator becomes an interlayer exciton insulator with finite exciton density $n_X = p_{mono} > 0$. Beyond a critical exciton density $n_x^m \sim 0.50 p_0$, the interlayer exciton insulator melts into a metallic phase. Within the interlayer exciton insulator states, exciton superfluidity might arise above a critical $n_X$, but further experimental probes will be needed to examine such behaviors.

Last, we compare the temperature dependence of the Mott insulator and correlated interlayer exciton insulator states. Figure 4 shows the 2s exciton transition probing the Mott insulator state (top panels, $E = 160$ mV/nm) and the correlated interlayer exciton insulator state (lower panels, $E = 0$ mV/nm) at $T = 30$ K (a), 60 K (b), 90 K (c), 120 K (d), and 150 K (e). The 2s exciton resonance intensity decreases monotonically as the temperature is increased. The 2s exciton resonance associated with the Mott insulator at $p_{moiré}/p_0 = 1$ can be observed up to $T = 150$ K (upper panels of Fig. 4). However, the 2s exciton resonance associated with the correlated insulator state at $p/p_0 = (p_{moiré} + p_{mono})/p_0 = 1$ disappears at a temperature above $T = 60$ K (lower panels of Fig. 4). It shows that the correlated interlayer exciton insulator state has a low phase transition temperature than the Mott insulator state.

In conclusion, we demonstrate a new correlated interlayer exciton insulator composed of electrons in a Mott insulator and holes in a band insulator using the double-layer heterostructure that combines strongly coupled $WS_2/WSe_2$ moiré bilayer and $WSe_2$ monolayer. The combination of flat bands from the moiré superlattice and strong interlayer interactions in the double layers provides an exciting new avenue for engineering novel correlated interlayer phases that are not possible in an isolated moiré superlattice. Many exotic quantum phases could arise in such systems: for example, if the Mott insulator is a spin-liquid, the correlated interlayer exciton insulator may carry new quantum excitations with fractionalized statistics[29,30].

**Methods**:

Heterostructure preparation for optical measurements:

All the two-dimensional flakes are first exfoliated from the bulk crystal on the $SiO_2$/Si substrate, and we stack them up by using a polypropylene carbon (PPC) based dry transfer technology[31]. The moiré bilayer is composed of $WS_2$/$WSe_2$ heterostructures with a near-zero or 60-degrees twist angle. The crystal orientations of these two flakes are determined optically using polarization-dependent second harmonic generation (SHG) measurements before the transfer process. The moiré bilayer and $WSe_2$ monolayer are separated by an ultrathin hexagonal boron nitride (hBN) layer with a thickness of approximately 1nm. These atomic double layers are contacted separately by few-layer graphite (FLG). The top and bottom gates are made of FLG, and two thicker hBN flakes serve as the top and bottom dielectric layers with dielectric constant $\varepsilon_{hBN} \sim 4.2 \pm 0.4$ (ref. (*13*)). The heterostructure is further capped by an hBN flake to ensure the cleanness of top graphite and aid in the device assembly. Finally, the whole stack is released onto a 90 nm $SiO_2$/Si substrate. Electrodes (5 nm Cr /100 nm Au) are defined by a standard photolithography system (Durham Magneto Optics, MicroWriter) and an e-beam deposition system. The top and bottom gate voltages are applied by Keithley 2400 or 2450 source meters. The $WSe_2$ monolayer, moiré bilayer, and heavily hole-doped Si are grounded during the measurements.

Optical measurements:

The optical measurements are performed in a cryostat with a temperature down to $T = 1.6$ K (Quantum Design, Opticool). We use diode lasers as the light source for reflection spectroscopy. The light is focused on the sample by a 20X Mitutoyo objective with ~2 μm beam size. The reflected light is collected by the same objective and dispersed by a spectrometer before reaching the camera. We take three spectra to get the reflection contrast spectrum: spectrum on the sample ($R_s$), spectrum without the sample ($R_{ref}$), and background spectrum ($R_{bkg}$). The reflection contrast ($-DR/R$) is calculated as $-(R_s-R_{ref})/(R_{ref}-R_{bkg})$. The noise level is ~ 0.1% in our measurements.

Photoluminescence measurements are performed using a 532 nm continuous laser source, which is spectrally filtered by a 650 nm low-pass filter. The excitation light is focused on the sample by a 20X Mitutoyo objective with ~2 μm beam size and then filtered out by a 700 nm long-pass filter. Photoluminescence is collected and analyzed with a monochromator and a camera. The excitation power is around 2 μW, and the integration time is 1 minute.

**Data availability**

The data that support the findings of this study are available from the corresponding author upon reasonable request.

**Competing interests:**
The authors declare that they have no competing interests.

**Acknowledgements**

This work was supported primarily by the US Department of Energy, Office of Science, Office of Basic Energy Sciences, Materials Sciences and Engineering Division under contract number DE-AC02-05-CH11231 (van der Waals heterostructures programme, KCWF16). The device fabrication was also supported by the US Army Research Office under MURI award W911NF-

17-1-0312. ECR acknowledges support from the Department of Defense through the National Defense Science & Engineering Graduate Fellowship (NDSEG) Program. ST acknowledges support from DOE-SC0020653, NSF CMMI 1933214, NSF mid-scale 1935994, NSF 1904716, NSF DMR 1552220, and DMR 1955889. KW and TT acknowledge support from the Elemental Strategy Initiative conducted by the MEXT, Japan, Grant Number PMXP0112101001, JSPS KAKENHI Grant Number JP20H00354, and the CREST(JPMJCR15F3), JST.

**Author contributions**

FW conceived the research. ZZ fabricated the device and performed most of the experimental measurements. ECR, DW, and WZ contributed to the optical measurement. ZZ and FW performed data analysis. ECR, DW, WZ, SW, MC, and AZ contributed to the fabrication of van der Waals heterostructures. MZ contributed to the theory. MS, KY, and ST grew $WSe_2$ and $WS_2$ crystals. KW and TT grew hBN crystals. All authors discussed the results and wrote the manuscript.

**References**

1. Mott, N. F. The transition to the metallic state. *Philos. Mag. J. Theor. Exp. Appl. Phys.* **6**, 287–309 (1961).

2. HALPERIN, B. I. & RICE, T. M. Possible Anomalies at a Semimetal-Semiconductor Transistion. *Rev. Mod. Phys.* **40**, 755–766 (1968).

3. Kuneš, J. Excitonic condensation in systems of strongly correlated electrons. *J. Phys. Condens. Matter* **27**, 333201 (2015).

4. Eisenstein, J. P. Exciton Condensation in Bilayer Quantum Hall Systems. *Annu. Rev. Condens. Matter Phys.* **5**, 159–181 (2014).

5. Liu, X., Watanabe, K., Taniguchi, T., Halperin, B. I. & Kim, P. Quantum Hall drag of exciton condensate in graphene. *Nat. Phys.* **13**, 746–750 (2017).

6. Li, J. I. A., Taniguchi, T., Watanabe, K., Hone, J. & Dean, C. R. Excitonic superfluid phase in double bilayer graphene. *Nat. Phys.* **13**, 751–755 (2017).

7. Liu, X. *et al.* Crossover between Strongly-coupled and Weakly-coupled Exciton Superfluids. *arXiv:2012.05916* (2020).

8. Cao, Y. *et al.* Correlated insulator behaviour at half-filling in magic-angle graphene superlattices. *Nature* **556**, 80–84 (2018).

9. Cao, Y. *et al.* Unconventional superconductivity in magic-angle graphene superlattices. *Nature* **556**, 43–50 (2018).

10. Chen, G. *et al.* Evidence of a gate-tunable Mott insulator in a trilayer graphene moiré superlattice. *Nat. Phys.* **15**, 237–241 (2019).

11. Chen, G. *et al.* Signatures of tunable superconductivity in a trilayer graphene moiré superlattice. *Nature* **572**, 215–219 (2019).

12. Tang, Y. *et al.* Simulation of Hubbard model physics in WSe$_2$/WS$_2$ moiré superlattices. *Nature* **579**, 353–358 (2020).

13. Regan, E. C. *et al.* Mott and generalized Wigner crystal states in WSe$_2$/WS$_2$ moiré superlattices. *Nature* **579**, 359–363 (2020).

14. Balents, L., Dean, C. R., Efetov, D. K. & Young, A. F. Superconductivity and strong correlations in moiré flat bands. *Nat. Phys.* **16**, 725–733 (2020).

15. Yankowitz, M. *et al.* Tuning superconductivity in twisted bilayer graphene. *Science* **363**, 1059–1064 (2019).

16. Lu, X. *et al.* Superconductors, orbital magnets and correlated states in magic-angle bilayer graphene. *Nature* **574**, 653–657 (2019).

17. Sharpe, A. L. *et al.* Emergent ferromagnetism near three-quarters filling in twisted bilayer graphene. *Science* **365**, 605–608 (2019).

18. Serlin, M. *et al.* Intrinsic quantized anomalous Hall effect in a moiré heterostructure. *Science* **367**, 900–903 (2020).

19. Chen, G. *et al.* Tunable correlated Chern insulator and ferromagnetism in a moiré superlattice. *Nature* **579**, 56–61 (2020).

20. Nuckolls, K. P. *et al.* Strongly correlated Chern insulators in magic-angle twisted bilayer graphene. *Nature* **588**, 610–615 (2020).

21. Seyler, K. L. *et al.* Signatures of moiré-trapped valley excitons in MoSe$_2$/WSe$_2$ heterobilayers. *Nature* **567**, 66–70 (2019).

22. Tran, K. *et al.* Evidence for moiré excitons in van der Waals heterostructures. *Nature* **567**, 71–75 (2019).

23. Jin, C. *et al.* Observation of moiré excitons in WSe$_2$/WS$_2$ heterostructure superlattices. *Nature* **567**, 76–80 (2019).

24. Alexeev, E. M. *et al.* Resonantly hybridized excitons in moiré superlattices in van der Waals heterostructures. *Nature* **567**, 81–86 (2019).

25. Xu, Y. *et al.* Correlated insulating states at fractional fillings of moiré superlattices. *Nature* **587**, 214–218 (2020).

26. Wang, G. *et al.* Colloquium: Excitons in atomically thin transition metal dichalcogenides. *Rev. Mod. Phys.* **90**, 021001 (2018).

27. Fogler, M. M., Butov, L. V. & Novoselov, K. S. High-temperature superfluidity with indirect excitons in van der Waals heterostructures. *Nat. Commun.* **5**, 4555 (2014).

28. Wu, F.-C., Xue, F. & MacDonald, A. H. Theory of two-dimensional spatially indirect equilibrium exciton condensates. *Phys. Rev. B* **92**, 165121 (2015).

29. Barkeshli, M., Nayak, C., Papić, Z., Young, A. & Zaletel, M. Topological Exciton Fermi Surfaces in Two-Component Fractional Quantized Hall Insulators. *Phys. Rev. Lett.* **121**, 026603 (2018).

30. Hu, Y., Venderbos, J. W. F. & Kane, C. L. Fractional Excitonic Insulator. *Phys. Rev. Lett.* **121**, 126601 (2018).

31. Wang, L. *et al.* One-Dimensional Electrical Contact to a Two-Dimensional Material. *Science* **342**, 614–617 (2013).

**Figure caption**:

**Figure 1 | Schematic of double layers composed of $WS_2/WSe_2$ moiré bilayer and $WSe_2$ monolayer. a,** A schematic of double-layer heterostructure with both the top and bottom gates. Both moiré bilayer and $WSe_2$ monolayer can be hole-doped in the system. **b,** Band alignment of the three TMDC layers in the double layers. The conduction band minimum is in the moiré superlattice $WS_2$ layer. Holes can be distributed in either the $WSe_2$ monolayer or the moiré superlattice $WSe_2$ layer, and the distribution depends on the applied vertical electric field $E$. **c,** Broadband reflection spectrum when both moiré bilayer and $WSe_2$ monolayer are charge neutral. The 1s and 2s exciton resonances of the $WSe_2$ monolayer and the moiré excitons of the $WS_2/WSe_2$ bilayer can be observed.

**Figure 2 | Correlated insulating states in double layers.** Two-dimensional color plots of derivative reflection contrast spectra and PL spectra in three different photon energy windows (row I, II, and III) at the $E$ field of 160 mV/nm (column **a**), 16 mV/nm (column **b**), 0 mV/nm (column **c**), and -16 mV/nm (column **d**). (**Row I**): Derivative spectra of the $WSe_2$ monolayer 2s exciton probes the correlated insulating state in the double-layer system. (**Row II**): $WSe_2$ monolayer 1s exciton and trion PL spectra probe the hole density in the $WSe_2$ monolayer. (**Row III**): IX PL spectra of the $WS_2/WSe_2$ moiré bilayer. The Mott insulator state in the moiré bilayer is characterized by an abrupt increase of the IX PL intensity and a blueshift of the IX energy. **a,** The monolayer $WSe_2$ 2s exciton resonances at $p/p_0$ = 1/3, 2/3, and 1 in (**aI**) correspond to the generalized Wigner crystal and Mott insulator states. The $WSe_2$ monolayer is charge neutral with no trion PL signal in (**aII**), and the Mott insulator state occurs at $p/p_0$ =1 in (**aIII**). **b-d,**

Prominent 2s exciton resonances are still observed at $p/p_0 = 1$ in **bI** to **dI**, indicating a correlated insulator state when the total doping is at one hole per moiré superlattice site. Trion PL starts to appear at $p/p_0 = 1$ in bII, and becomes increasingly stronger in **cII** and **dII**. It shows that the correlated insulator has partial hole doping in the WSe$_2$ monolayer. At the same time, significantly higher total hole densities ($p/p_0 > 1$) are required to realize a filled moiré superlattice (i.e., $p_{moiré}/p_0 = 1$) in **bIII** to **dIII**, confirming that holes are now distributed in both the WSe$_2$ monolayer and the WS$_2$/WSe$_2$ moiré bilayer at $p/p_0 = 1$. The correlated insulators in **bI** to **dI** represent new interlayer exciton insulator states in the double-layer system.

**Figure 3 | Phase diagram of the correlated interlayer exciton insulator. a,** WSe$_2$ monolayer integrated trion area as a function of total doping $p/p_0$ and vertical electric field $E$. The black dashed line, determined by the emergence of the WSe$_2$ monolayer trion PL, separates region I (charge neutral WSe$_2$ monolayer) and region II (hole-doped WSe$_2$ monolayer). **b,** Derivative of IX PL intensity with respect to total hole doping $dI_{IX}/d(p/p_0)$. The Mott insulator state at $p_{moiré}/p_0 = 1$, characterized by a maximum $dI_{IX}/d(p/p_0)$ (green dashed line), coincides with the vertical line defining $p/p_0 = 1$ in region I. In region II, a higher total hole density $p/p_0$ is required to sustain the Mott insulator, confirming that an increasingly larger portion of holes is doped into the WSe$_2$ monolayer. **c,** The WSe$_2$ monolayer 2s exciton signal, which shows a stable correlated insulator state at $p/p_0 = 1$, extends region I to region II. The correlated insulator in region I, where $p_{mono} = 0$, corresponds to the Mott insulator in the moiré bilayer. The correlated insulator in region II represents a new correlated interlayer exciton insulator at a combined hole density of $p/p_0 = (p_{moiré} + p_{mono})/p_0 = 1$. **d,** Hole distribution of the Mott insulator at point A of **(c)**. **e,** Hole

distribution of the interlayer correlated insulator with an effective "one hole per moiré lattice site" at point B of **(c)**. The holes in the WSe$_2$ monolayer will avoid positions below the WS$_2$/WSe$_2$ moiré lattice site that is occupied by a hole due to the strong interlayer Coulomb interaction. **f,** Particle-hole transformation of the doped Mott insulator state. Electrons doped into the Mott insulator in the moiré layer spontaneously bind the holes doped into the WSe$_2$ monolayer to form the interlayer exciton insulator. **g,** Different correlated phases in the double layers as a function of $p_{mono}$ at a fixed $p/p_0 = 1$.

**Figure 4 | Temperature dependence of the Mott insulator and correlated interlayer exciton insulator. a-e,** Correlated insulating states at $T = 30$ K **(a)**, 60 K **(b)**, 90 K **(c)**, 120 K **(d)**, and 150 K **(e)**. The upper (lower) panels plot the derivative spectra of 2s exciton resonance in the WSe$_2$ monolayer at $E = 160$ mV/nm ($E = 0$ V/nm). The Mott insulator state, revealed by the 2s resonance at $p/p_0 = p_{moiré}/p_0 = 1$ in the upper panel, persists up to 150 K. The correlated interlayer exciton insulator, revealed by the 2s resonance at $p/p_0 = (p_{moiré} + p_{mono})/p_0 = 1$ in the lower panel state, is stable up to 60 K.

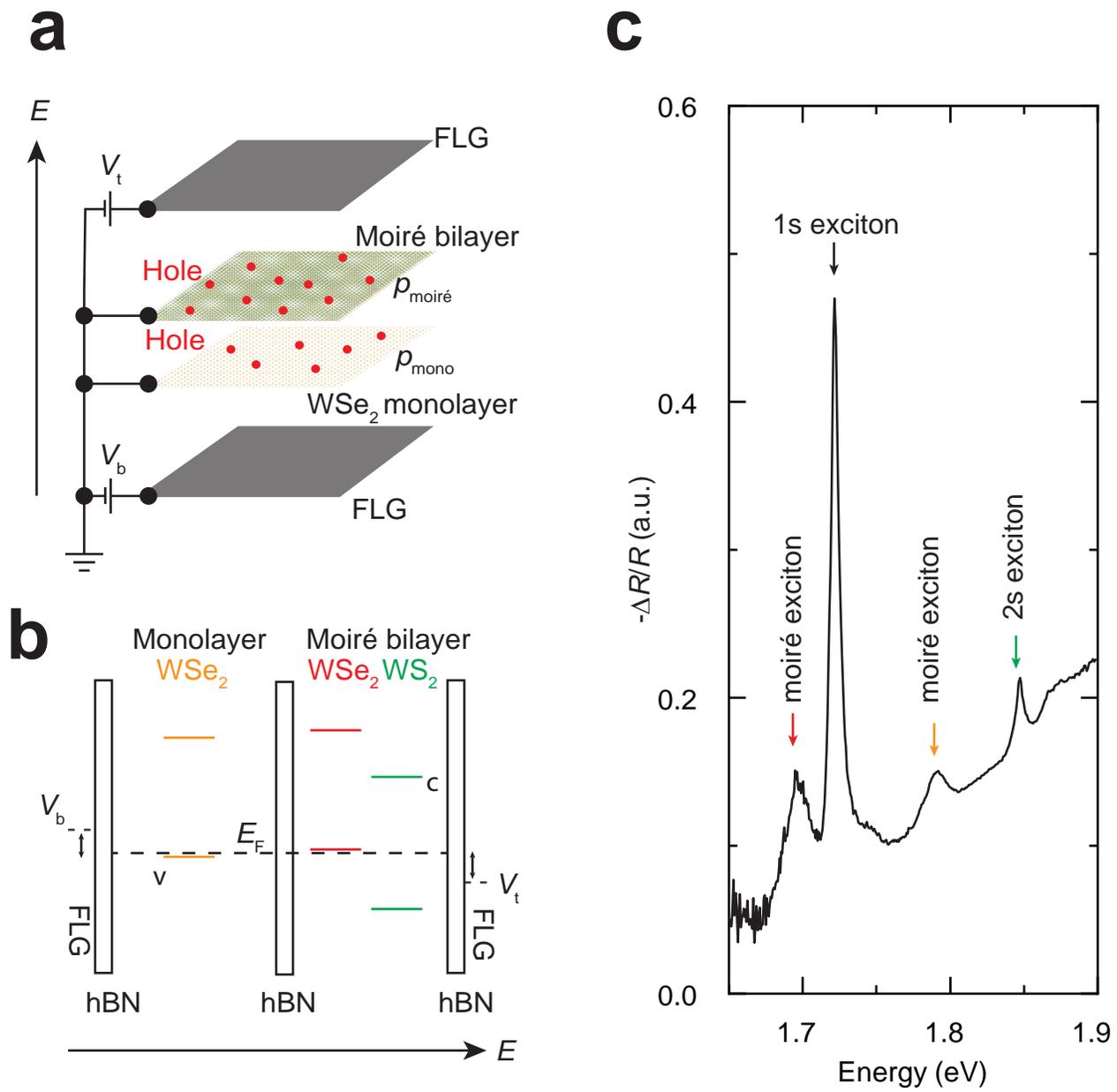

Fig. 1

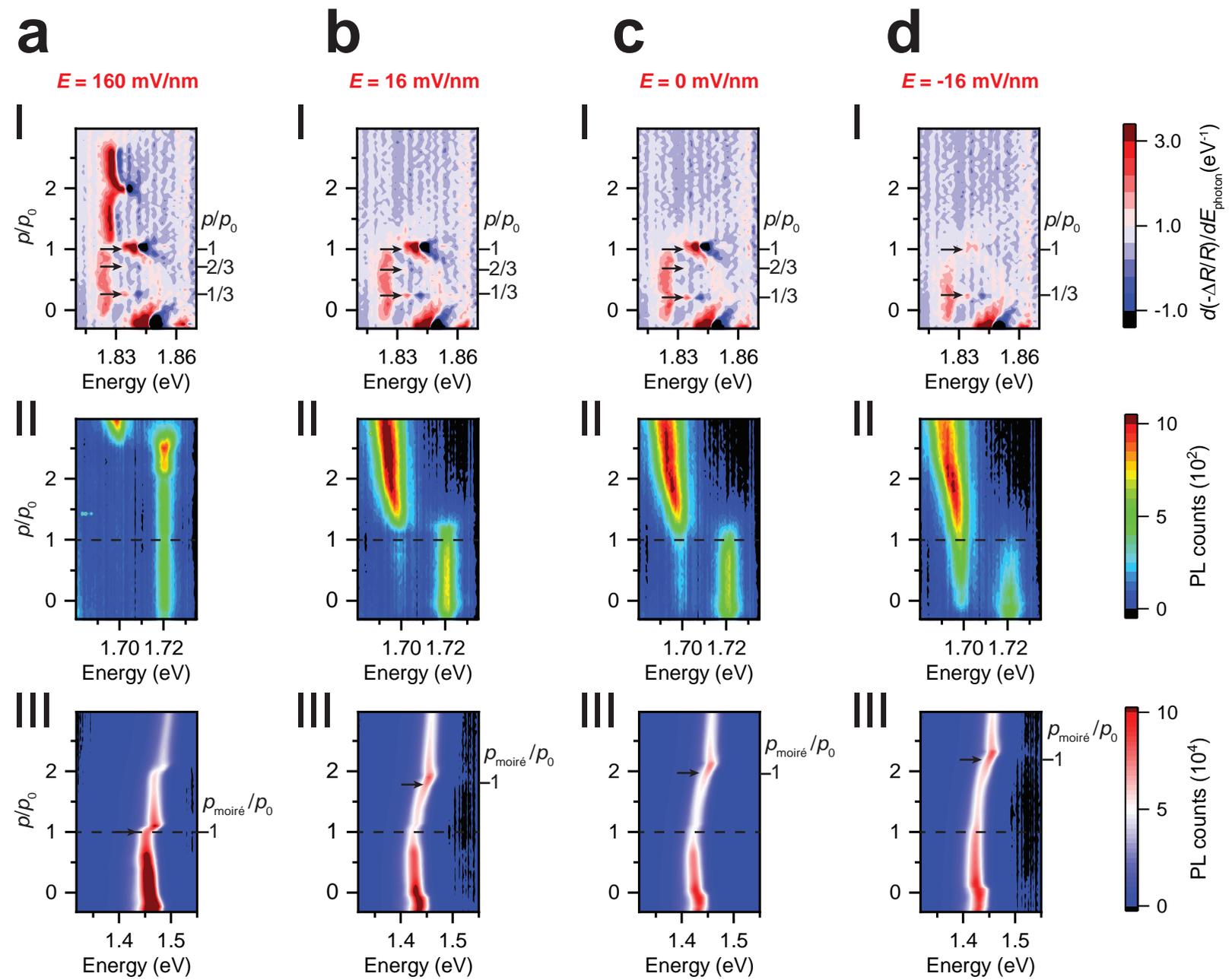

Fig. 2

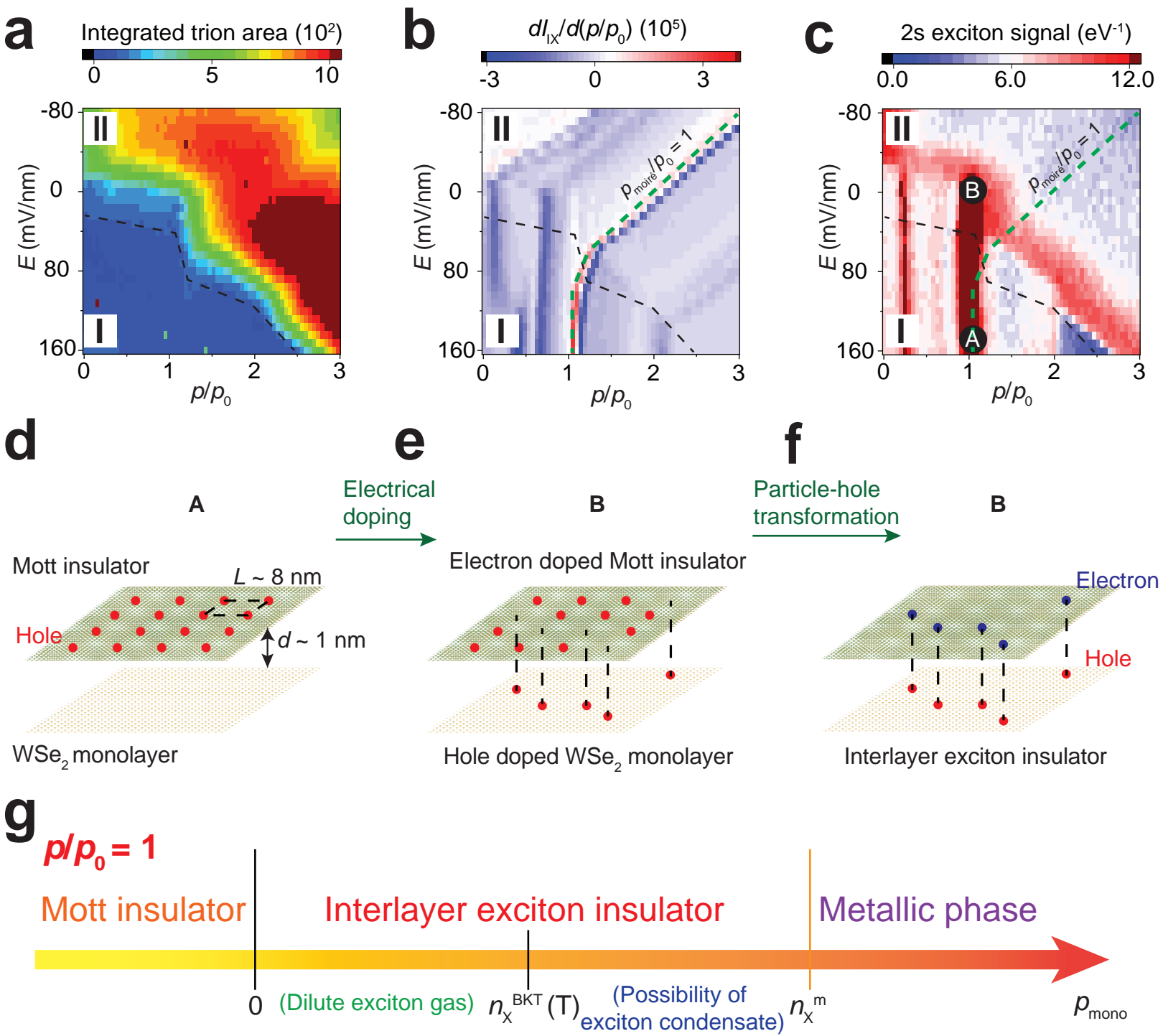

Fig. 3

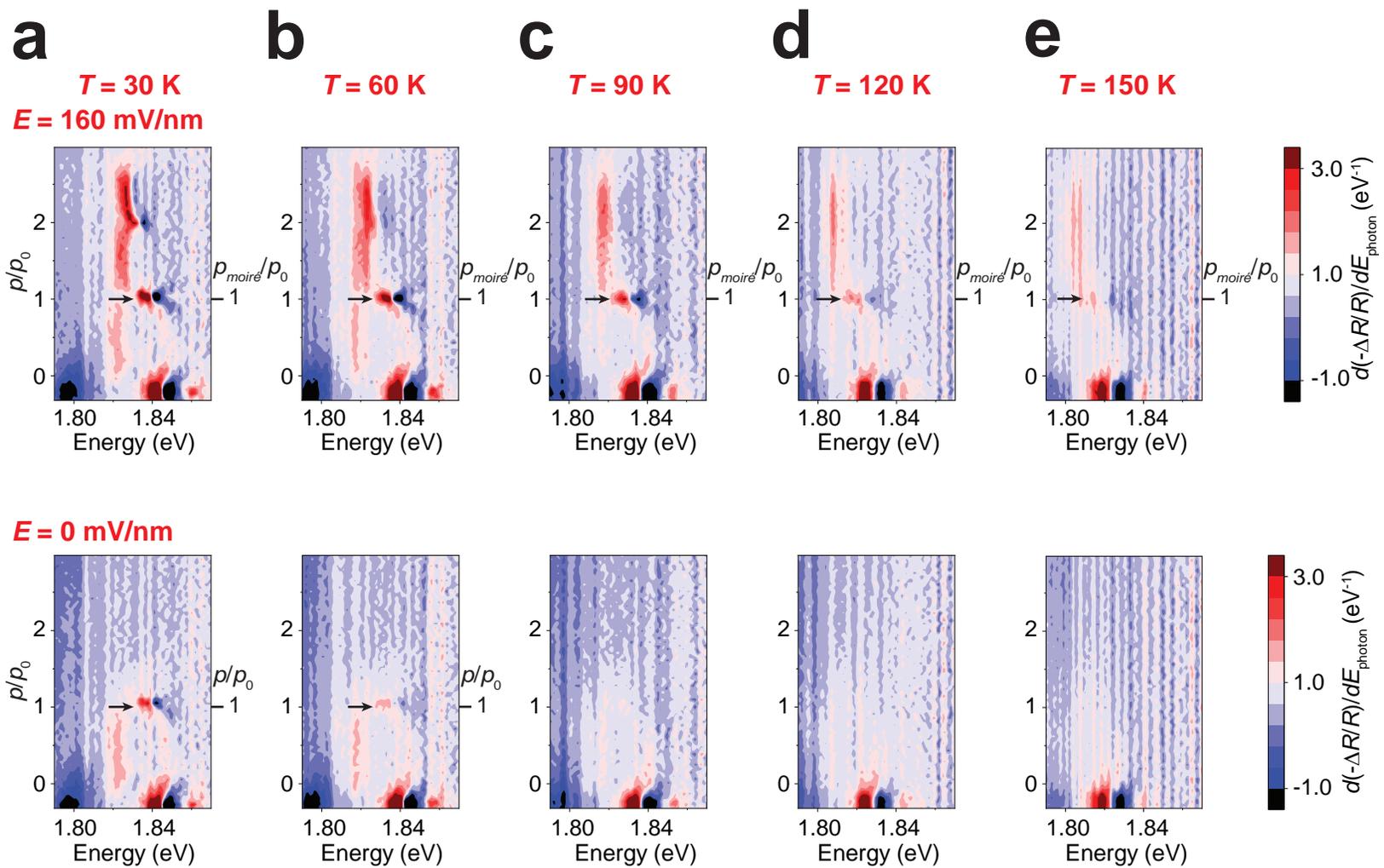

Fig. 4

Supplementary Information for

Correlated interlayer exciton insulator in double layers of monolayer $WSe_2$ and moiré $WS_2/WSe_2$

Zuocheng Zhang, Emma C. Regan, Danqing Wang, Wenyu Zhao, Shaoxin Wang, Mohammed Sayyad, Kentaro Yumigeta, Kenji Watanabe, Takashi Taniguchi, Sefaattin Tongay, Michael Crommie, Alex Zettl, Michael P. Zaletel, Feng Wang*

*Correspondence to: fengwang76@berkeley.edu

1. Device information
2. Determination of hole density at one hole per moiré lattice site
3. Phase diagram of the correlated interlayer exciton insulator in device II and device III
4. Estimation of the critical interlayer exciton density

1. Device information

Figure S1 depicts three side-view schematics of our double-layer heterostructures device I (Fig. S1a), device II (Fig. S1b), and device III (Fig. S1c). In all three devices, the moiré bilayer and WSe₂ monolayer are separated by an ultrathin hBN with a thickness of approximately 1 nm thickness. Dual top ($V_t$) and bottom ($V_b$) gates are included to independently control the total hole density $p = -\frac{1}{e}[\frac{\varepsilon_{hBN}\varepsilon_0}{d_t}(V_t - V_{t0}) + \frac{\varepsilon_{hBN}\varepsilon_0}{d_b}(V_b - V_{b0})]$ and the vertical electric field $E = (\frac{1}{d_b}V_b - \frac{1}{d_t}V_t)$, where $V_{t0}$ and $V_{b0}$ are the onset gate voltages to inject holes into the double layers, where $d_b$ ($d_t$) is the thickness of the bottom (top) hBN dielectric layer, $\varepsilon_{hBN}$ is the dielectric constant of hBN with $\varepsilon_{hBN} \approx 4.2$ (ref. 1), $\varepsilon_0$ is the vacuum permittivity, $e$ is the electron charge. The stacking order of device III is opposite to that in device I and device II. The moiré bilayer in device III is below the WSe₂ monolayer. To be consistent with the schematic shown in Fig. 1a in the main text, we use $V_t$ to denote the gate close to the moiré bilayer and $V_b$ to denote the gate close to the WSe₂ monolayer in all three devices.

In device I, we use relatively thick hBN gate dielectric layers with the top hBN thickness of approximately 50 nm and the bottom hBN thickness of approximately 50 nm, as determined by atomic force measurements (AFM). The electrical contacts to the transition metal dichalcogenide (TMDC) flakes and the top/bottom gates are made of FLG. The stack is further capped with an approximately 65 nm thick hBN layer. The middle panel of Fig. S1a displays an optical microscopy image of this device. The green (red) dashed line outlines the WS₂ (WSe₂) layer in the moiré heterostructure, the yellow dashed line outlines the monolayer WSe₂, the black dashed line outlines the top FLG gate, and the white dashed line outlines the bottom FLG gate. These three flakes are aligned as demonstrated by the SHG signal on WSe₂ in the moiré bilayer, WS₂ in the moiré bilayer, and WSe$_2$ monolayer (lower panel in Fig. S1a). The moiré bilayer in device I has a near 60-degrees twist angle.

In device II, we also use relatively thick hBN gate dielectric layers with the top hBN thickness of approximately 80 nm and the bottom hBN thickness of approximately 40 nm, as determined by AFM. The electrical contacts to the TMDCs and the top/bottom gates are made of FLG. The stack is further capped with an approximately 55 nm thick hBN layer. The middle panel of Fig. S1b displays an optical microscopy image of this device. The green (red) dashed line outlines the WS$_2$ (WSe$_2$) layer in the moiré heterostructure, the yellow dashed line outlines the monolayer WSe$_2$, the black dashed line outlines the top FLG gate, and the white dashed line outlines the bottom FLG gate. The twist angle between the moiré bilayer and WSe$_2$ monolayer is around 25 degrees, as determined by the SHG signal (lower panel of Fig. S1b). The WS$_2$/WSe$_2$ moiré bilayer has a near zero-twist-angle as the moiré bilayer shows a relatively enhanced SHG signal compared to that on the monolayer WSe$_2$ in the moiré bilayer. The moiré bilayer in device II has a near zero-degree twist angle.

In device III, we use relatively thin hBN gate dielectric layers with the top hBN thickness of approximately 9 nm and the bottom hBN thickness of approximately 15 nm, as estimated by optical contrast under the microscope. The electrical contacts to the TMDCs and the top/bottom gates are made of FLG. The stack is further capped with an approximately 8 nm thick hBN layer. The middle panel of Fig. S1c displays an optical microscopy image of this device. The green (red) dashed line outlines the WS$_2$ (WSe$_2$) layer in the moiré heterostructure, the yellow dashed line outlines the monolayer WSe$_2$, the black dashed line outlines the top FLG gate, and white dashed line outlines the bottom FLG gate. The solid lines correspond to a few-layer TMDCs that are attached to the monolayers. These three layers are also aligned, as determined by the SHG measurements in the lower panel of Fig. S1c. The moiré bilayer in device III has a near zero-degree twist angle.

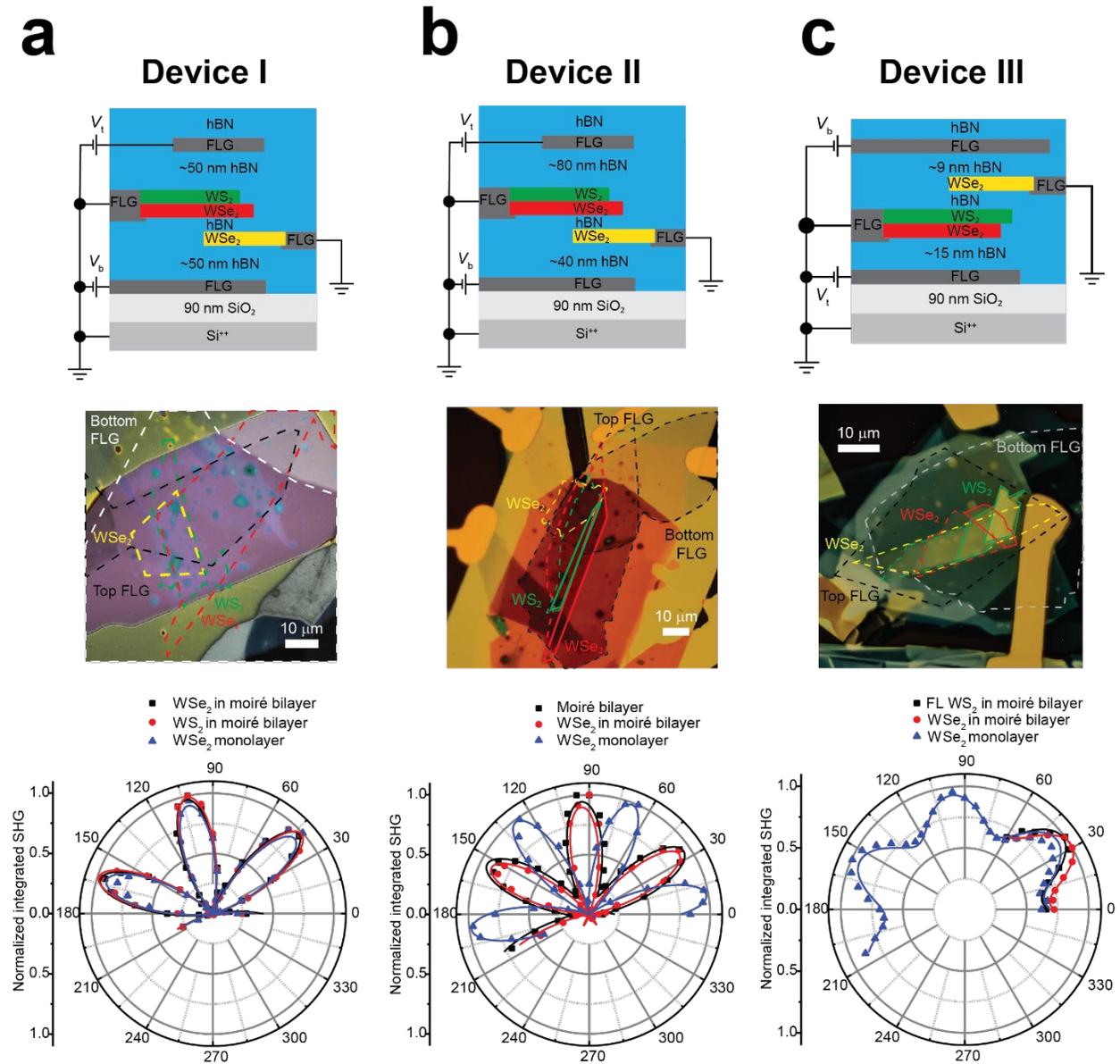

**Fig. S1. Device information. a-c,** Three side-view schematics, optical microscope image, and SHG results of device I (a), device II (b), and device III (c). From top to bottom are the side-view schematic (upper panels), optical microscope image (middle panels), and SHG results (lower panels). To be consistent with the schematic shown in Fig. 1a in the main text, we use $V_t$ to denote the gate close to the moiré bilayer and $V_b$ to denote the gate close to the WSe$_2$ monolayer in device III since the stacking order is opposite to the other two devices.

## 2. Determination of hole density at one hole per moiré lattice site

We use the geometry capacitance to determine the hole density at one hole per moiré lattice site concentration $p_0$. We assign the filling factor $p/p_0$ = 2, 1, 2/3, and 1/3 to the enhanced resonance of the 2s exciton transition and $p/p_0 = 0$ to the point that the double layers start to be doped in Fig. S2a. The total hole concentration $p$, calculated as $p = -\frac{1}{e}[\frac{\varepsilon_{hBN}\varepsilon_0}{d_t}(V_t - V_{t0}) + \frac{\varepsilon_{hBN}\varepsilon_0}{d_b}(V_b - V_{b0})]$, is then plotted as a function of the five filling factors in Fig. S2b at vertical electric field $E = 160$ V/nm. The linear fit of these four points gives the $p_0 = (1.97 \pm 0.10) \times 10^{12}/cm^2$. Here, the error bar reflects how well the fit is.

The uncertainty of the hBN dielectric constant, thickness and the top gate voltage width of correlated states together determine the error bar in estimating the hole density at one hole per moiré lattice site. The hBN dielectric constant has a 10% experimental uncertainty (ref. [1]), and thus the $p_0 = (1.97 \pm 0.20) \times 10^{12}/cm^2$.

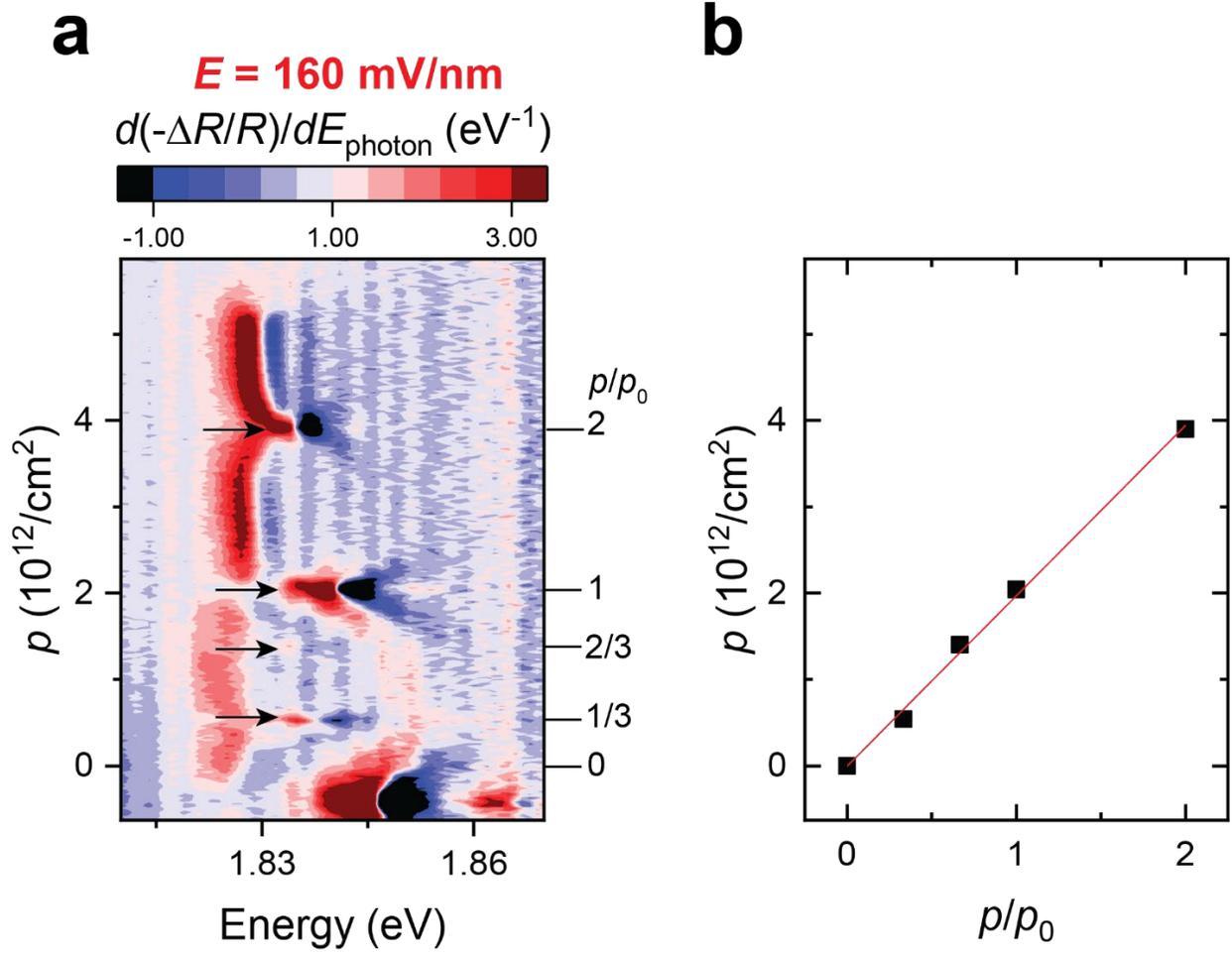

**Fig. S2. Determination of hole density at one hole per moiré lattice site. a,** The two-dimensional color plot of derivative reflection contrast spectra at $E = 160$ mV/nm. We assign the filling factor $p/p_0 = 2, 1, 2/3$, and $1/3$ to the enhanced resonance of the 2s exciton transition and $p/p_0 = 0$ to the point that the double layers start to be doped. **b,** The corresponding hole density $p$ at five filling factors. The linear fit to these five points gives the $p_0 = (1.97 \pm 0.10) \times 10^{12}/\text{cm}^2$. Here, the error bar reflects how well the fit is. As the hBN dielectric constant has a 10% experimental uncertainty, the hole density at one hole per moiré lattice site is $p_0 = (1.97 \pm 0.20) \times 10^{12}/\text{cm}^2$

## 3. Phase diagram of the correlated interlayer exciton insulator in device II and device III

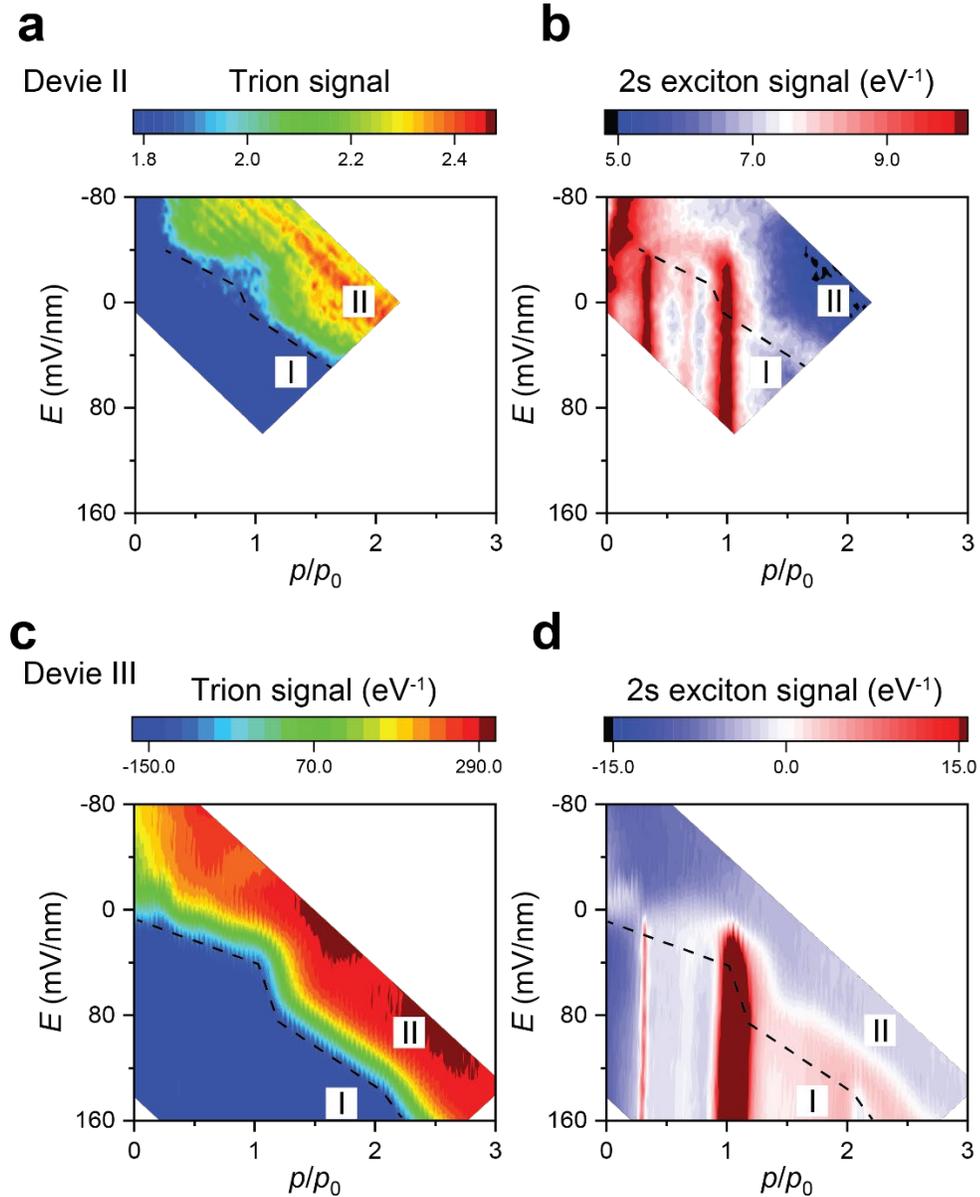

**Fig. S3. Phase diagram of the correlated interlayer exciton insulator in device II and device III. a** and **b,** WSe$_2$ monolayer trion area in the reflection contrast spectra (a) and WSe$_2$ monolayer 2s exciton area in the derivative reflection contrast spectra (b) as a function of total hole doping concentration $p/p_0$ and vertical electric field $E$ in device II. The black dashed line, determined by the emergence of the trion signal in (a), separates region I and region II. WSe$_2$

monolayer is charge neutral in region I ($p_{mono} = 0$) and has finite hole doping in region II ($p_{mono} > 0$). The strongest 2s resonances at $p/p_0 = 1$ in the region I of (b) defines the Mott insulator state of the moiré bilayer. This Mott insulator state becomes a correlated interlayer exciton insulator as the resonance extends to region II of (b) at the combined hole density $p/p_0 = (p_{moiré} + p_{mono}) = 1$, and eventually disappears at the sufficiently high negative vertical electric field. **c** and **d,** WSe$_2$ monolayer trion area (c) and WSe$_2$ monolayer 2s exciton area (d) in the derivative reflection contrast spectra as a function of total hole doping concentration $p/p_0$ and vertical electric field $E$ in device III. Here to emphasize the oscillator strength of hole-type trion and the 2s exciton, we use the first derivative of reflection contrast spectra with respect to photon energy for both trion and 2s exciton signal. The black dashed line, determined by the emergence of the trion signal in (c), separates region I and region II. WSe$_2$ monolayer is charge neutral in region I ($p_{mono} = 0$) and has finite hole doping in region II ($p_{mono} > 0$). The strongest 2s resonances in region I of (c) defines the Mott insulator state of the moiré bilayer. This Mott insulator state also becomes an interlayer exciton insulator as the resonance extends into region II of (b) with the combined hole density $p/p_0 = (p_{moiré} + p_{mono}) = 1$, and eventually disappears at the negative vertical electric field.

## 4. Estimation of the critical interlayer exciton density

We replot Fig. 3a to 3c, in the main text to Fig. S4a to S4c in the supplementary information to estimate the critical interlayer exciton density $n_X$, beyond which the correlated interlayer exciton becomes metallic. The critical point is labeled as point X in Fig. S4a to S4c.

We estimate the hole density in the WSe$_2$ monolayer ($p_{mono}$) at the point X through the trion PL intensity. Different double-layer configurations with the same trion PL intensity will have approximately the same $p_{mono}$ because the trion PL emission is dominated by holes in the WSe$_2$ monolayer only. The black dashed line in Fig. S4a denotes states with the same trion PL intensity, and therefore similar $p_{mono}$, as the X point.

The green dashed line in Fig. S4a to S4c, corresponds to the Mott insulator in the moiré bilayer ($p_{moiré}/p_0 = 1$), which is determined by the maximum of $dI_{IX}/d(p/p_0)$ in Fig. S4b. The carrier density in the WSe$_2$ monolayer $p_{mono}$ along this green dashed line can be calculated as $p_{mono} = p - p_{moiré}$.

The intersection of the black dashed line and the green dashed line in Fig. S4a is around $p/p_0 = 1.5$. The hole density in the WSe$_2$ monolayer $p_{mono}$ is $p_{mono} = p - p_{moiré} = 1.5p_0 - p_0 = 0.5p_0$ at the intersection. Point X should have a similar $p_{mono}$ because it has the same trion intensity. Consequently, we estimate the critical interlayer exciton density at X point to be $n_X = p_{mono} \sim 0.5p_0$.

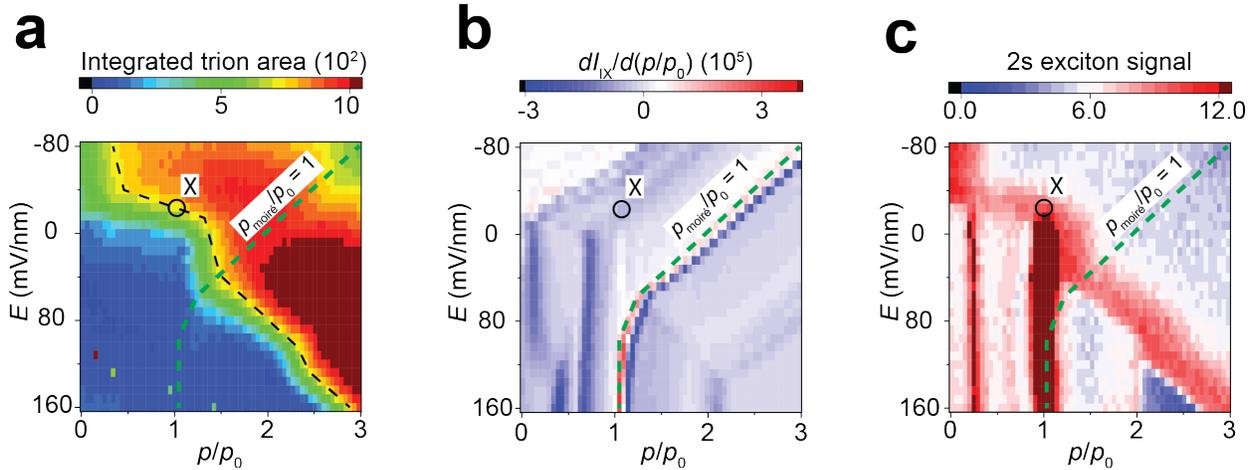

**Fig. S4. Estimation of critical interlayer exciton density. a-c,** WSe$_2$ monolayer integrated trion area (a), the derivative of IX PL intensity with respect to total hole doping (b), and WSe$_2$ monolayer 2s exciton signal (c) as a function of total doping $p/p_0$ and vertical electric field $E$.

The same figures are plotted in Fig. 3a to 3c, in the main text. The critical point is labeled as point X, where the correlated interlayer exciton insulator starts to melt, as shown in (c). The black dashed line in (a) denotes states with the same trion PL intensity, and thus similar $p_{mono}$, as the X point. The green dashed line in (a) corresponds to the Mott insulator in the moiré bilayer ($p_{moiré}/p_0 = 1$), which is determined by the maximum of $dI_{IX}/d(p/p_0)$ in (b). The carrier density in the WSe$_2$ monolayer $p_{mono}$ along this green dashed line can be calculated as $p_{mono} = p - p_{moiré}$. The intersection of the black dashed line and the green dashed line is around $p/p_0 = 1.5$. The hole density in the WSe$_2$ monolayer $p_{mono}$ is $p_{mono} = p - p_{moiré} = 1.5p_0 - p_0 = 0.5p_0$ at the intersection. Point X should have a similar $p_{mono}$ because it has the same trion intensity. Therefore, the critical interlayer exciton density at point X is $n_X = p_{mono} \sim 0.5p_0$.

**References**:

1. Regan, E. C. *et al.* Mott and generalized Wigner crystal states in WSe$_2$/WS$_2$ moiré superlattices. *Nature* **579**, 359–363 (2020).